\documentclass[a4paper]{llncs}
\usepackage{graphicx}
\usepackage{url}
\usepackage{hyperref}
\usepackage{amsmath}
\usepackage{amssymb}
\usepackage{enumerate}
\newcommand{\SA}{G}%\mathcal{A}}
\newcommand{\SW}{M}%\mathcal{W}}
\newcommand{\IR}{I}%\mathcal{W}}

\def\KK{\mathbb{K}}
\def\BB{\underline{\mathfrak{B}}}

\begin{document}

\title{Concept Stability for Constructing Taxonomies of Web-site Users}%
\author{Sergei O. Kuznetsov and Dmitrii I. Ignatov}
\institute{State University Higher  School of Economics (HSE) and
All-Russia
Institute for Scientific and Technical Information (VINITI),\\
Moscow, Russia}

\maketitle
% ----------------------------------------------------------------
\begin{abstract}

Owners of a web-site are often interested in analysis of groups of
users of their site. Information on these groups can help
optimizing the structure and contents of the site. In this paper
we use an approach based on formal concepts for constructing
taxonomies of user groups. For decreasing the huge amount of
concepts that arise in applications, we employ stability index of
a concept, which describes how a group given by a concept extent
differs from other such groups. We analyze resulting taxonomies of
user groups for three target websites.

\end{abstract}

% ----------------------------------------------------------------

\subsection*{Problem Statement and Domain Models}

Owners of a web-site are often interested in analyzing groups of
users of their site. Information on these groups can help to
optimize the structure and contents of the site. For example,
interaction with members of each group may be organized in a
special manner. In this paper we use an approach based on formal
concepts \cite{GW99} for constructing taxonomies of user groups.

For our experiments we have chosen four target websites: the site
of the State University Higher School of Economics (www.hse.ru),
an e-shop of household equipment, the site of a large bank, and
the site of a car e-shop (the names of the last three sites cannot
be disclosed due to legal agreements).

Users of these sites are described by attributes that correspond
to other sites, either external (from three groups of sites:
finance, media, education) or internal (web-pages of the site).
More precisely, initial ``external" data consists of user records
each containing the user id, the time when the user first entered
this site, the time of his/her last visit, and the total number of
sessions during the period under consideration. An ``internal" user
record, on the other hand, is simply a list of pages within the
target website visited by a particular user.

By ``external" and ``internal" taxonomies we mean (parts of) concept
lattices for contexts with  either ``external" or ``internal"
attributes. For example, the external context has the form

$$K_e= (U,S_e, I_e),$$
where $U$  is the set of all users of the target site,  $S_e$  is the
set of all sites from a sample (not including the target one), the
incidence relation $I_e$ is given by all pairs $(u,s)\colon  u\in U, s\in
S_e$, such that user $u$ visited site $s$. Analogously, the internal
context is of the form $K_i=(U,S_i, I_i)$, where $S_i$ is the set
of all own pages of the target site.

A concept of this context is a pair $(A, B)$ such that $A$ is a
group of users that visited together all other sites from $B$.

\section*{Formal framework}

Before proceeding, we briefly recall the FCA terminology
\cite{GW99}. Given a \emph{(formal) context} $\KK = (\SA, \SW,
\IR)$, where $\SA$ is called a set of \emph{objects}, $\SW$ is
called a set of \emph{attributes}, and the binary relation ${\IR
\subseteq \SA\times \SW}$ specifies which objects have which
attribute, the derivation operators $(\cdot)^I$ are defined for $A
\subseteq \SA$ and $B \subseteq \SW$ as follows:
\[A^I=\{m \in \SW \mid \forall g \in A: g \IR m\};\]
\[B^I=\{g \in \SA \mid \forall m \in B: g \IR m\}.\]
Put differently, $A^I$ is the set of attributes common to all
objects of $A$ and $B^I$ is the set of objects sharing all
attributes of $B$.

If this does not result in ambiguity, $(\cdot)'$ is used instead
of $(\cdot)^I$. The double application of $(\cdot)'$ is a closure
operator, i.e., $(\cdot)''$ is extensive, idempotent, and
monotonous. Therefore, sets $A''$ and $B''$ are said to be
\emph{closed}.

A \emph{(formal) concept} of the context $(G, M, I)$ is a pair
$(A, B)$, where $A \subseteq G$, $B \subseteq M$, $A = B'$, and $B
= A'$. In this case, we also have $A = A''$ and $B = B''$. The set
$A$ is called the \emph{extent} and $B$ is called the
\emph{intent} of the concept $(A, B)$. In categorical terms,
$(A,B)$ is equivalently defined by its objects $A$ or its
attributes $B$.

A concept $(A, B)$ is a \emph{subconcept} of $(C, D)$ if $A
\subseteq C$ (equivalently, $D \subseteq B$). In this case, $(C,
D)$ is called a \emph{superconcept} of $(A, B)$. We write $(A, B)
\leq (C, D)$ and define the relations $\geq$, $<$, and $>$ as
usual. If $(A, B) < (C, D)$ and there is no $(E, F)$ such that
$(A, B) < (E, F) < (C, D)$, then $(A, B)$ is a \emph{lower
neighbor} of $(C, D)$ and $(C, D)$ is an \emph{upper neighbor} of
$(A, B)$; notation: $(A, B) \prec (C, D)$ and $(C, D) \succ (A,
B)$.

The set of all concepts ordered by $\leq$ forms a lattice, which
is denoted by $\BB(\KK)$ and called the \emph{concept lattice} of
the context $\KK.$ The relation $\prec$ defines edges in the
\emph{covering graph} of $\BB(\KK)$.

\subsection*{Data and Their Preprocessing}

We received ``external" data with the following  information for each
user-site pair: ({\bf user id, time of the first visit, time of the last visit,
total number of sessions during the period}).
``Internal" data have almost the same format with an additional field
{\bf url page}, which corresponds to a particular visited page of the target site.

Information was gathered from about 10000 sites of Russian
internet (domain .ru). In describing users in terms of sites they visites we
had to tackle the problem of dimensionality, since concept
lattices can be very large (exponential in the worst case) in terms
of attributes. To reduce
the size of input data we used the following techniques.

For each user we selected only those
sites that were visited by more than a certain number of times
during the observation period. This gave us information about
permament interests of particular users. Each target site was
considered in terms of sites of three groups: newspaper sites, financial sites, and
educational sites. Some pages can
be merged (as attributes) according to (implicit) domain ontology.
For example, if users of a bank site have personal pages, it is
reasonable to fuse all these pages by calling the resulting attribute ``a
personal web-page". A certain observation period can be chosen;
usually we took a one-month period.

However, even for large reduction of input size, concept lattices
can be very large. For example, a context of size 4125$\times$225 gave
rise to a lattice with 57 329 concepts.

\subsection*{Using Stability for Selecting Interesting Subsets of Concepts}

To choose  interesting groups of users we employed stability
index of a concept defined in \cite{k90,Kuz03} and considered in
\cite{ROK06} (in slightly different form) as a tool for
constructing taxonomies. On one hand, stability index shows the independence
of an intent on particular objects of extent (which may appear or not appear in the context depending
on random factors). On the other hand, stability index of a concept
shows how much extent of a concept is different from similar smaller extents
(if this difference is very small, then its doubtful that extent refers to a
``stable category''). For detailed motivation of staibility indices see~\cite{k90,Kuz03,ROK06}.
\bigskip

{\bf Definition.} Let $K = (G, M, I)$ be a formal context and $(A,
B)$ be a formal concept of $K$. The stability index $\sigma$ of
$(A, B)$ is defined as follows:

$$\sigma(A,B) = \frac{|\{C\subseteq A | C'=A \}|}{2^{|A|}}.$$
\bigskip

Obviously, $0 \leq \sigma (A, B) \leq 1$. The stability index of a
concept indicates how much the concept intent depends on
particular objects of the extent. A stable intent (with stability
index close to 1) is probably ``real" even if the description of
some objects is ``noisy". In application to our data, the stability
index shows how likely we are to still observe a common group of
interests if we ignore several users. Apart from being noise-resistance,
a stable group does not collapse (e.g., merge with a different
group, split into several independent subgroups) when a few
members of the group stop attending the target sites.

In our experiments we used ConceptExplorer~\cite{ce} for computing and visualizing lattices
and their parts.
We compared results of taking most stable concepts (with stability
index exceeding a threshold) with taking an ``iceberg" of a concept
lattice (order filter of a lattice containing all concepts with
extents larger than a threshold). The results look correlated, but
nevertheless, substantially different. The set of stable extents
contained very important, but not large groups of users.

In Figs. 1, 2  we present parts of a concept lattice for the site
www.hse.ru described by ``external" attributes which were taken to be Russian
internet newspapers visited by users of  www.hse.ru during one month more than 20 times.
Fig. 1  presents an iceberg
with 25 concepts having largest extent. Many of the concepts correspond to
newspapers that are in the middle of political spectrum, read ``by everybody" and thus,
not very interesting in characteizing social groups.

\includegraphics[scale=0.7]{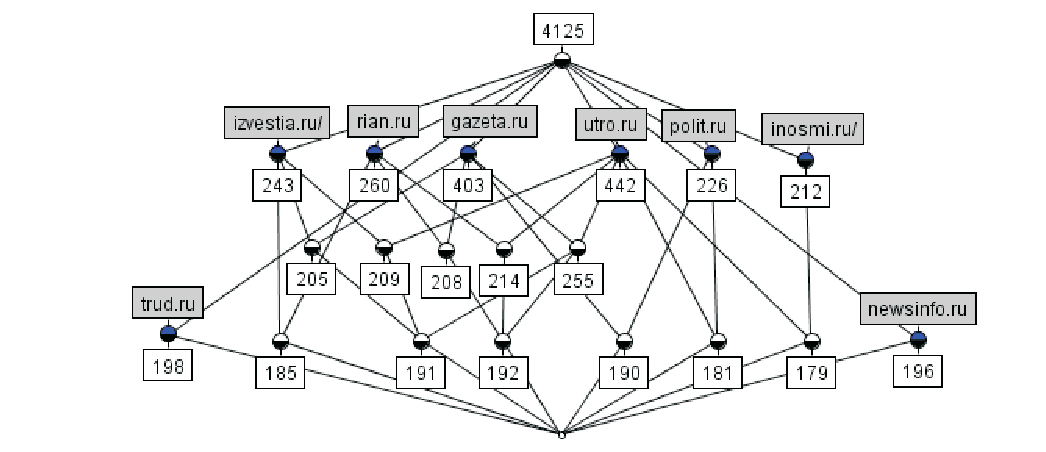}\\
Fig. 1. Iceberg with 25 concepts

\vspace{0.5cm}

Fig. 2  presents an ordered set of
25 concepts having largest stability index. As compared to the iceberg,
this part of the concept latice contains several sociologically important groups
such as readers of AIF (``yellow press"), Cosmopolitain, Expert (high professional
analytical surveys) etc.

\includegraphics[scale=0.7]{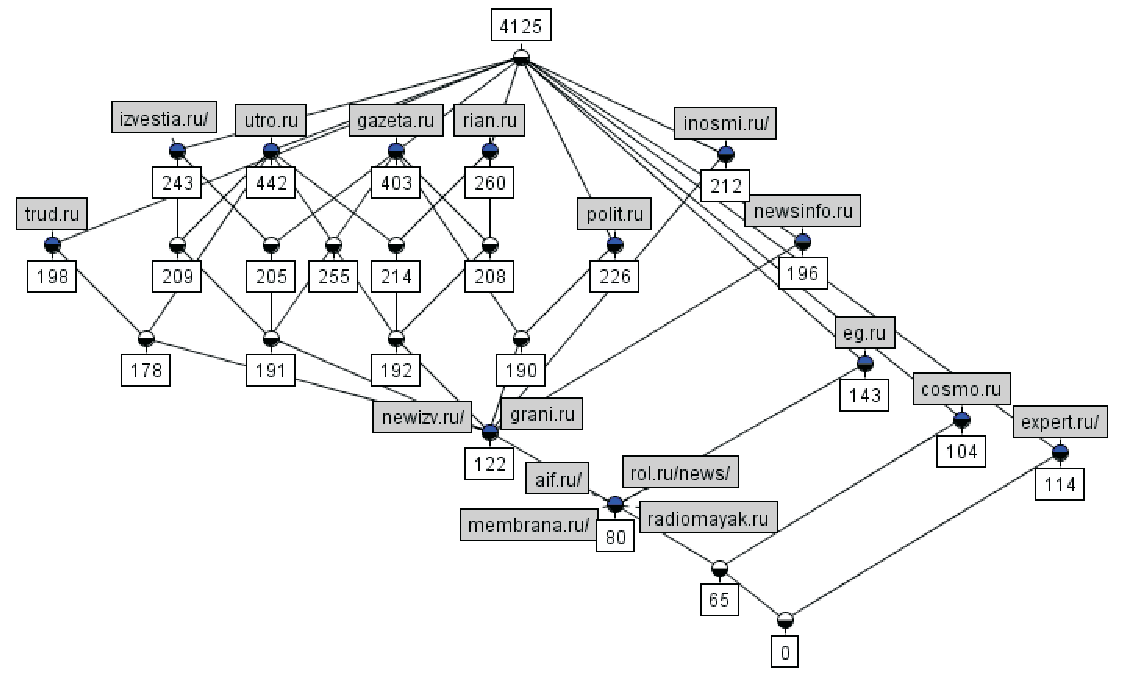}\\
Fig. 2. Ordered set of 25 concepts with largest stability

% ----------------------------------------------------------------

\end{document}